\pdfoutput=1
\newcommand*{\ATLASLATEXPATH}{latex/}
\documentclass[cernpreprint,texlive=2016,txfonts,UKenglish]{\ATLASLATEXPATH atlasdoc}
\usepackage[backend=bibtex]{\ATLASLATEXPATH atlaspackage}
\usepackage{\ATLASLATEXPATH atlasbiblatex}
\usepackage[jetetmiss]{\ATLASLATEXPATH atlasphysics}
\graphicspath{{logos/}{figures/}}

\usepackage{paper_draft-defs}
\usepackage{longtable}

\AtlasTitle{Search for new phenomena in high-mass final states with a photon and a jet from \pp collisions at 
\sqs = 13~TeV with the ATLAS detector}

\author{The ATLAS Collaboration}

\AtlasRefCode{EXOT-2016-26}

\PreprintIdNumber{CERN-EP-2017-148}

\AtlasJournal{EPJC}

\AtlasAbstract{
A search is performed for new phenomena in events having a photon with high transverse momentum and a jet collected in 
\lumi of proton--proton collisions at a centre-of-mass energy of \sqs = 13~TeV recorded with the ATLAS detector at 
the Large Hadron Collider. 
The invariant mass distribution of the leading photon and jet is examined to look for the resonant production 
of new particles or the presence of new high-mass states beyond the Standard Model. 
No significant deviation from the background-only hypothesis is observed 
and cross-section limits for generic Gaussian-shaped resonances are extracted. 
Excited quarks hypothesized in quark compositeness models and high-mass states predicted in quantum 
black hole models with extra dimensions are also examined in the analysis. 
The observed data exclude, at 95\% confidence level, the mass range below \mqstarObsLimit~TeV for excited quarks 
and \mADDObsLimit~TeV (\mRSObsLimit~TeV) 
for quantum black holes in the Arkani-Hamed--Dimopoulos--Dvali (Randall--Sundrum) model with six (one) extra dimensions.
}

\hypersetup{pdftitle={ATLAS document},pdfauthor={The ATLAS Collaboration}}

\begin{document}

\maketitle

\section{Introduction}
\label{sec:intro}
This paper reports a search for new phenomena in events with a photon and a jet produced from proton--proton (\pp) collisions at \sqs = 13~TeV,
collected with the ATLAS detector at the Large Hadron Collider (LHC). 
Prompt photons in association with jets are copiously produced at the LHC, mainly through 
quark--gluon
scattering 
($qg \to q\gamma$). The $\gamma$ + jet(s) final state provides a sensitive 
probe for a 
class of 
phenomena beyond the Standard Model (SM) that could manifest themselves in the high invariant mass (\mgj) region 
of the $\gamma$ + jet system. 
The search is performed by looking for
localized excesses of events in the \mgj distribution
with respect to the SM prediction.
Two classes of benchmark signal models are considered. 

The first class of benchmark models is based on a generic Gaussian-shaped mass distribution with different values of its
mean and standard deviation. This provides a generic interpretation for the presence of signals with different Gaussian widths,
ranging from a resonance with a width similar to the reconstructed \mgj resolution of $\sim2$\% to wide resonances 
with a width up to 15\%. 
The second class of benchmark models is based on signals beyond the SM 
that are implemented in
Monte Carlo (MC) simulation and 
appear as broad peaks in the \mgj spectrum.
This paper considers two scenarios for physics beyond the SM:
quarks as composite particles and extra spatial dimensions.
In the first case, if quarks are composed of more fundamental constituents bound together by some 
unknown interaction, new effects should appear depending on the value of the compositeness scale 
$\Lambda$. In particular, if $\Lambda$ is sufficiently smaller than the centre-of-mass energy, 
excited quark (\qstar) states may be produced in high-energy \pp collisions at the LHC ~\cite{Baur:1987ga,PhysRevD42815,
Bhattacharya:2009p2168}. 
The \qstar production at the LHC could result in 
a resonant peak at the mass of the \qstar (\mqstar) in the \mgj distribution 
if the \qstar can decay into a photon and a quark ($qg \to \qstar \to q\gamma$).
In the present search, only the SM gauge interactions are considered for \qstar production.
In the second scenario, the existence of extra spatial dimensions (EDs) is assumed to provide a solution to the hierarchy 
problem~\cite{Weinberg:1978ym,Veltman:1980mj,Smith:1981yi}. 
Certain types of ED models predict the fundamental Planck scale \mstar in the $4+n$ 
dimensions ($n$ being the number of extra spatial dimensions) to be at the TeV scale, and thus accessible 
in \pp collisions at \sqs = 13~TeV at the LHC. In such a TeV-scale \mstar scenario of the extra dimensions, quantum black holes (QBHs) 
may be produced at the LHC as a continuum above the threshold mass (\mth) and then 
decay into a small number of final-state particles including photon--quark/gluon pairs before they are able
to thermalize
\cite{Dimopoulos:2001hw,Giddings:2001bu,Gingrich:2009hj,Calmet:2008dg}. 
In this case a broad resonance-like structure could be observed just above \mth on top of the SM \mgj distribution.
The \mth value for QBH production is taken to be equal to \mstar while the maximum allowed QBH mass is set 
to either $3\mstar$ or the LHC \pp centre-of-mass energy of 13~TeV, whichever is smaller. 
The upper bound on the mass ensures that the QBH production is far from the ``thermal'' regime, where the classical 
description of the black hole and its decay into high-multiplicity final states should be used.
In this paper, the extra-dimensions model proposed by 
Arkani-Hamed, Dimopoulous and Dvali (ADD)~\cite{ArkaniHamed:1998rs} with $n=6$ flat EDs and the one by 
Randall and Sundrum (RS1)~\cite{Randall:1999ee} with $n=1$ warped ED 
are considered. 

The ATLAS and CMS experiments at the LHC have performed searches for excited quarks in the \gj final state 
using \pp collision data recorded at \sqs = 7~TeV~\cite{EXOT-2011-04}, 8~TeV~\cite{EXOT-2012-22,CMS-EXO-13-003} 
and 13~TeV~\cite{EXOT-2015-12}. In the ATLAS searches, limits for generic Gaussian-shaped resonances were 
obtained at 7, 8 and 13~TeV while a limit for QBHs in the ADD model ($n=6$) was first obtained at 8~TeV. 
The ATLAS search at 13~TeV with data taken in 2015 was further extended to constrain QBHs 
in the RS1 model ($n=1$).
No significant excess of events was observed in any of these searches, and 
the lower mass limits of 4.4~TeV for the \qstar and 6.2 (3.8)~TeV for QBHs in the ADD (RS1) model
were set, currently representing the most stringent limits for the decay into a photon and a jet.
For a Gaussian-shaped resonance a cross-section upper limit of 0.8 (1.0)~fb at \sqs = 13~TeV was obtained,
for example, for a mass of 5~TeV and a width of 2\% (15\%).

The dijet resonance searches at ATLAS~\cite{EXOT-2015-02,EXOT-2016-21} and CMS~\cite{CMS-EXO-16-032} 
using \pp collisions 
at \sqs = 13~TeV also set limits 
on the production cross-sections of excited quarks and QBHs.
The search in the \gj final state presented here complements the dijet results and 
provides an independent check for the presence of these signals in different decay channels.

This paper presents the search based on the full 2015 and 2016 data set recorded with the ATLAS detector, corresponding to
\lumi of \pp collisions at \sqs = 13~TeV. The analysis strategy is unchanged from the one reported in Ref.~\cite{EXOT-2015-12}, focusing 
on the region where the $\gamma$ + jet system has a high invariant mass.
 
The paper is organized as follows. In Section~\ref{sec:detector} a brief description of the ATLAS detector is given. Section~\ref{sec:data_sim} summarizes the data and simulation samples used in this study. The event selection is discussed in 
Section~\ref{sec:selection}. 
The signal and background modelling are presented in Section~\ref{sec:stat_analysis} together with the signal search and 
limit-setting strategies. Finally the results are discussed in Section~\ref{sec:result} and the conclusions are given 
in Section~\ref{sec:conclusion}.

\section{ATLAS detector}
\label{sec:detector}
\newcommand{\AtlasCoordFootnote}{%
ATLAS uses a right-handed coordinate system with its origin at the nominal interaction point (IP)
in the centre of the detector and the $z$-axis along the beam pipe.
The $x$-axis points from the IP to the centre of the LHC ring,
and the $y$-axis points upwards.
Cylindrical coordinates $(r,\phi)$ are used in the transverse plane, 
$\phi$ being the azimuthal angle around the $z$-axis.
The pseudorapidity is defined in terms of the polar angle $\theta$ as $\eta = -\ln \tan(\theta/2)$.
Angular distance is measured in units of $\Delta R \equiv \sqrt{(\Delta\eta)^{2} + (\Delta\phi)^{2}}$.}

The ATLAS detector at the LHC is a multi-purpose, forward-backward symmetric detector\footnote{\AtlasCoordFootnote} 
with almost full solid angle coverage, and is described in 
detail elsewhere~\cite{PERF-2007-01,ATLAS-TDR-19}. 
Most relevant for this analysis are the inner detector (ID) and the calorimeter system composed of 
electromagnetic (EM) and hadronic calorimeters. The ID consists of a silicon pixel detector, 
a silicon microstrip tracker and a transition radiation tracker, all immersed in a \SI{2}{\tesla} axial magnetic field,
and provides charged-particle tracking in the range $|\eta|<2.5$.
The electromagnetic calorimeter is a lead/liquid-argon (LAr) sampling calorimeter with accordion geometry.
The calorimeter is divided into a barrel section covering $|\eta|<1.475$ and two endcap sections covering $1.375<|\eta|<3.2$. 
For $|\eta|<2.5$ it is divided into three layers in depth,  which are finely segmented in $\eta$ and $\phi$. 
In the region $|\eta|<1.8$, an additional
thin LAr presampler layer is used to correct for fluctuations in the energy losses in the material upstream 
of the calorimeters.
The hadronic calorimeter is a sampling calorimeter composed of steel/scintillator tiles 
in the central region ($|\eta|<1.7$), while 
copper/LAr modules are used in the endcap ($1.5<|\eta|<3.2$) regions. The forward region ($3.1<|\eta|<4.9$) is instrumented 
with copper/LAr and tungsten/LAr calorimeter modules optimized for electromagnetic and hadronic measurements,
respectively.
Surrounding the calorimeters is a muon spectrometer that includes three air-core superconducting toroidal magnets
and multiple types of tracking chambers,
providing precision tracking for muons within $|\eta|<2.7$ and trigger capability within $|\eta|<2.4$. 

A dedicated two-level trigger system is used for the online event selection~\cite{TRIG-2016-01}. 
Events are selected using a first-level trigger implemented in custom electronics, which reduces the event rate to 
a design value of \SI{100}{\kilo\hertz} using a subset of the detector information.
This is followed by a software-based trigger that 
reduces the accepted event rate to \SI{1}{\kilo\hertz} on average by refining the first-level trigger selection.

\section{Data and Monte Carlo simulations}
\label{sec:data_sim}
The data sample used in this analysis was collected from \pp collisions in the 2015--2016 LHC run at a centre-of-mass energy of 13~TeV, 
and corresponds to an integrated luminosity of \lumierr. The uncertainty was derived, following a methodology
similar to that detailed in Ref.~\cite{DAPR-2013-01}, from a preliminary calibration of the luminosity scale using $x$--$y$ beam-separation scans
performed in August 2015 and May 2016. The data are required to satisfy a number of quality criteria ensuring that the relevant detectors
were operational while the data were recorded. 

Monte Carlo samples of simulated events are used to study the background modelling for the dominant \gj processes, 
to optimize the selection criteria and to evaluate the acceptance and selection efficiencies for the signals considered in the search.
Events from SM processes containing a photon with associated
jets were simulated using
the \SHERPA~2.1.1~\cite{Gleisberg:2008ta} event generator, 
requiring a photon transverse energy $\eT^{\gamma}$ above 70~GeV at the generator level.
The matrix elements were calculated with up to four final state partons at leading order (\LO) in quantum chromodynamics (QCD) and merged
with the parton shower ~\cite{Schumann:2007mg} using the ME+PS@LO prescription~\cite{Hoeche:2009rj}. 
The CT10~\cite{Lai:2010vv} parton distribution 
function (PDF) set was used in conjunction with dedicated parton shower tuning developed by the \SHERPA authors.
A second sample of SM \gj events was generated using the LO \PYTHIA~8.186~\cite{Sjostrand:2014zea} event generator with
the LO NNPDF~2.3 PDFs and the A14 tuning of the underlying-event parameters ~\cite{ATL-PHYS-PUB-2014-021}. 
The \PYTHIA simulation includes leading-order $\gamma$ + jet events from both the direct processes (the hard subprocesses
$qg \rightarrow q\gamma$ and $q\bar{q} \rightarrow g\gamma$) and bremsstrahlung photons in QCD dijet events.  
To estimate the systematic uncertainty associated with the background modelling, a large sample of events was generated 
with the next-to-leading-order (NLO) \jetphox~v1.3.1\_2~\cite{Catani:2002ny} program.
Events were generated at parton level for both the
direct and fragmentation photon contributions using the NLO photon fragmentation 
functions~\cite{Bourhis2001} and the NLO NNPDF~2.3~\cite{Ball:2012cx} PDFs, and 
setting the renormalization, factorization and fragmentation scales to the photon $\eT^{\gamma}$.
Jets of partons are reconstructed using the \akt algorithm~\cite{Cacciari:2008gp,Cacciari:2011ma} 
with a radius parameter of $R=0.4$. The generated photon is 
required to be isolated by ensuring that the total transverse energy of partons inside a cone of size $\DeltaR=0.4$ 
around the photon is smaller than 7.07~GeV + $0.03\times\eT^{\gamma}$,
equivalent \footnote{The parton-level isolation requirement takes into account the correlation between reconstruction-level
isolation energies and particle-level isolation energies, as a proxy for the parton-level isolation,
as evaluated using \gj events simulated by \PYTHIA~8.186.}
to the photon selection for the data described in Section~\ref{sec:selection}. 

Samples of excited quark events were produced using \PYTHIA~8.186
with 
the LO NNPDF~2.3 PDFs and 
the A14 set
of tuned parameters for the underlying event. 
The Standard Model gauge interactions and the magnetic-transition type couplings \cite{Baur:1987ga,PhysRevD42815,Bhattacharya:2009p2168}
to gauge bosons were considered in the production processes of the excited states of the first-generation quarks ($u^*$, $d^*$) 
with degenerate masses. The compositeness scale $\Lambda$ was taken to be equal to the mass \mqstar of the excited quark,
and the coefficients $f_\mathrm{s}$, $f$ and $f'$ 
of magnetic-transition type couplings to the respective SU(3), SU(2) and U(1) 
gauge bosons were chosen to be unity. The \qstar samples were generated with \mqstar values between 0.5 and 6.0~TeV in steps of 
0.5~TeV.

The QBH samples were generated using the QBH~2.02~\cite{Gingrich:2009da} event generator with the CTEQ6L1~\cite{Pumplin:2002vw} 
PDF set and 
\PYTHIA 8.186 for the parton shower and underlying event tuned with the A14 parameter set. 
The \mth values were chosen to vary between 3.0 (1.0) and 9.0 (7.0)~TeV in steps of 0.5~TeV for the QBH signals in the ADD (RS1)
model. 
All the $qg$, $\bar{q}g$, $gg$ and $q\bar{q}$ processes were included in the QBH signal production while 
only final states with
a photon and a quark/gluon were considered for the decay. All six quark flavours were included together with their anti-quark 
counterparts in both the production and decay processes.

Apart from the sample generated with \jetphox which is a parton-level calculator, all the simulated samples include the 
effects of multiple \pp interactions in the same and neighbouring 
bunch crossings (pile-up) and were processed through the ATLAS detector simulation ~\cite{SOFT-2010-01} based on \textsc{Geant4}
~\cite{Agostinelli:2002hh}. 
Pile-up
effects were emulated by overlaying simulated minimum-bias events from \PYTHIA~8.186, 
generated with the A2 tune~\cite{ATLAS:2012uec} for the
underlying event and the MSTW2008LO PDF set ~\cite{Martin:2009iq}. 
The number of overlaid minimum-bias events was adjusted to match the observed data.
All the MC samples except for the \jetphox sample were reconstructed with the same software as that used for collision data.

\section{Event selection}
\label{sec:selection}
Photons are reconstructed from clusters of energy deposits in the EM calorimeter as described in Ref.~\cite{Aaboud:2016yuq}. 
A photon candidate is classified depending on 
whether the EM cluster is associated with a conversion track candidate reconstructed in the ID. If no ID track is matched, 
the candidate is considered as an unconverted photon. 
If the EM cluster is matched to either a conversion vertex formed from two tracks constrained to originate 
from a massless particle or a single track with its first hit after the innermost layer of the pixel detector, the candidate is considered
to be a converted photon. Both the converted and unconverted photon candidates are used in the analysis.
The energy of each photon candidate is corrected using MC simulation and data as described in 
Ref.~\cite{PERF-2013-05}. The EM energy clusters are calibrated separately for converted and unconverted photons, 
based on their properties including the longitudinal shower development. The energy scale and resolution of the 
photon candidates after the MC-based calibration are further adjusted based on a correction derived using $Z \to \ee$ 
events from data and MC simulation,
respectively. 
Photon candidates are required to have $\etgamma>25$~GeV and $|\etagamma|<2.37$ 
and satisfy the ``tight'' identification 
criteria defined in Ref.~\cite{Aaboud:2016yuq}. 
Photons are identified 
based on the profile of the energy 
deposits in the first two layers of the EM calorimeter and the energy leakage into the hadronic calorimeter. To further reduce the 
contamination from \pizero $\rightarrow \gamma\gamma$ or other neutral hadrons decaying into photons, the photon candidates are 
required to be isolated 
from other energy deposits in an event. 
The calorimeter isolation variable 
\etiso
is defined as the sum of the $E_{\mathrm{T}}$ of all positive-energy topological clusters~\cite{PERF-2014-07}
reconstructed within a cone of $\DeltaR = 0.4$ around the photon direction excluding the energy deposits in an area of size 
$\Delta \eta \times \Delta \phi =0.125 \times 0.175$ 
centred on the photon cluster.
The photon energy expected outside the excluded area is subtracted from the isolation energy while the contributions from
pile-up and the underlying event are subtracted event by event~\cite{STDM-2010-08}.
The photon candidates are required to have 
$\etisogammaprime=\etiso-0.022\times\etgamma$
less than 2.45~GeV.
This \etgamma-dependent selection requirement is used to guarantee an efficiency greater than 90\% 
for signal photons in the \etgamma range relevant for this analysis.
The efficiency for the signal photon selection varies from ($90\pm1$)\% to ($83\pm1$)\% for signal events
with masses from 1 to 6~TeV. The dependency on the signal mass is mainly from the efficiency of the 
tight identification requirement while the isolation selection efficiency is approximately ($99\pm1$)\%
over the full mass range.

Jets are reconstructed from topological clusters calibrated at the electromagnetic scale using the \akt algorithm
with a radius parameter $R=0.4$. The jets are calibrated to the hadronic energy scale by applying corrections derived from
MC simulation and in situ measurements of relative jet response obtained from $Z$+jets, $\gamma$+jets and multijet events 
at \sqs = 13~TeV~\cite{PERF-2014-03, PERF-2016-04, ATLAS-CONF-2015-002}. 
Jets from pile-up interactions are suppressed by applying the jet vertex tagger~\cite{ATLAS-CONF-2014-018}, using 
information about tracks associated with the hard-scatter and pile-up vertices, to jets with $\ptjet<60$~GeV 
and $|\Etajet|<2.4$. 
 In order
to remove jets due to calorimeter noise or non-collision backgrounds, events containing at least one jet failing 
to satisfy the loose quality criteria defined in Ref.~\cite{ATLAS-CONF-2015-029} are discarded. Jets passing all the requirements and 
with $\ptjet>20$~GeV and $|\Etajet|<4.5$ are considered in the rest of the analysis.
Since a photon is also reconstructed as a jet, jet candidates in a cone of  $\DeltaR = 0.4$ around a photon
are not considered.

This analysis selects events based on a single-photon trigger requiring
at least one photon
candidate with $\et^{\gamma}>140$~GeV which satisfies at least loose identification
conditions~\cite{Aaboud:2016yuq}
based on the shower shape in the second sampling layer of the EM calorimeter and the energy leakage into the hadronic calorimeter.
Selected events are required to contain at least one primary vertex with two or more tracks 
with $\pT>400$~MeV. 
The kinematic requirements for the highest-\eT photon in the events are tightened to $\etgamma>150$~GeV and 
$|\etagamma|<1.37$. The \etgamma requirement is used to select events with nearly full trigger efficiency 
while the \etagamma requirement is imposed 
to enhance the signal-to-background ratio.
Moreover, an event is rejected if there is any jet with $\ptjet>30$~GeV within $\DeltaR<0.8$ around the photon. Finally, 
the highest-\pT jet in the event is required to have $\ptjet>60$~GeV and the pseudorapidity difference between the photon and 
the jet ($\detagj \equiv |\etagamma-\Etajet|$) must be less than 1.6
to enhance signals over the \gj background,                                                                                                                                   
which typically has a large \detagj value. 
The \gj system is formed from the highest-\eT photon and the highest-\pt jet
in the event.
After applying all the selection requirements, $6.34\times10^{5}$ events  with 
an invariant mass (\mgj) of the selected \gj system  
greater than 500 GeV remain in the data sample.

\section{Statistical analysis}
\label{sec:stat_analysis}
The data are examined for the presence of a significant deviation from the SM prediction using a test 
statistic based on a profile likelihood ratio ~\cite{Cowan:2010js}. Limits on the visible cross-section for generic 
Gaussian-shaped signals and limits on the cross-section times branching ratio for specific benchmark models are 
computed using the \CLs prescription ~\cite{Read:2002hq}. 
The details of the signal and background modelling used for the likelihood function construction 
are discussed in Sections~\ref{sec:signal_model} and \ref{sec:background_model}  while a summary of the 
statistical procedures used to establish 
the presence of a signal or set limits on the production cross-sections for new phenomena is given in 
Section~\ref{sec:fitting}.

\subsection{Signal modelling}
\label{sec:signal_model}

The signal model is built starting from the probability density function (pdf), \fsig, of the \mgj distribution 
at the reconstruction level. For a Gaussian-shaped resonance with mass \mg, the \mgj pdf is modelled by a normalized 
Gaussian distribution with the mean located at $\mgj = \mg$. The standard deviation of the Gaussian distribution is 
chosen to be 2\%, 7\% or 15\% of \mg, where 2\% approximately corresponds to the effect of the detector resolution 
on the reconstruction of the photon--jet invariant mass.
For the \qstar and QBH signals, the \mgj pdfs are created from the normalized reconstructed \mgj distributions after 
applying the selection requirements described in Section~\ref{sec:selection} using the simulated MC events, and a 
kernel density estimation technique~\cite{Cranmer:2000du} is applied to smooth the distributions.
The signal pdfs for intermediate mass points at which signal events were not generated are obtained from the simulated 
signal samples by using a moment-morphing method~\cite{Baak:2014fta}.
The signal template for the \qstar and QBH signals is then constructed as $\fsig \times (\sigma \cdot B \cdot A \cdot 
\varepsilon) \times \Lint$, where the \fsignom is scaled by the product of the cross-section times branching ratio to 
a photon and a quark or gluon (\sigmabr), acceptance ($A$), selection efficiency ($\varepsilon$) and the integrated 
luminosity (\Lint) for the data sample.
The product of the acceptance times 
efficiency ($A \cdot \varepsilon$) is found to be about 50\% for all the \qstar and QBH models, varying only by a few 
percent with \mqstar or \mth. This dependence is accounted for in the model by interpolating between the generated 
mass points using a third-order spline. 
The signal template for the Gaussian-shaped resonance is taken to be the \mgj pdf.
For the \qstar and QBH signals, limits are set on \sigmabr after correcting 
for the acceptance and efficiency $A \cdot \varepsilon$ of the selection criteria.

Experimental uncertainties in the signal yield arise from uncertainties in the luminosity ($\pm 3.2\%$),
photon identification efficiency ($\pm 2\%$),
trigger efficiency ($\pm 1\%$) and pile-up dependence ($\pm 1\%$). The impact of the uncertainties in the photon isolation 
efficiency, photon and jet energy scales and resolutions is negligible. A 1\% uncertainty in the signal yield is included 
to account for the statistical error in the acceptance and selection efficiency estimates due to the limited size of the MC signal  
samples.
The impact of the PDF uncertainties on the signal acceptance is found to be negligible compared to the other uncertainties.
The photon and jet energy resolution uncertainties ($\pm 2\%$ of the mass) are accounted for as a variation of the width for 
the Gaussian-shaped signals. The impact of the resolution uncertainty on intrinsically large width signals is found to be 
negligible and thus not included in the signal models for the \qstar and QBH. 
The typical difference between the peaks of the reconstructed and generator-level \mgj distributions
for the excited-quark signals is well below 1\%.

A summary of systematic uncertainties in the signal yield and shape included in the statistical analysis 
is given in Table~\ref{tab:systematics}.

\begin{table}[htbp]
  \centering
  \caption{Summary of systematic uncertainties in the signal event yield and shape included in the fit model. 
  The signal mass resolution uncertainty affects the generic Gaussian signal shape,  
while the other uncertainties affect the event yield.}
  \label{tab:systematics}
    \begin{tabular}{lccc}
    \toprule
    Uncertainty & \qstar and QBH & Generic Gaussian \\
    \midrule
    Signal mass resolution & N/A & $\pm 2\% \cdot m_{\mathrm{G}}$\\
    Photon identification & $\pm 2\%$ &  N/A \\
    Trigger efficiency & $\pm 1\%$ & N/A \\
    Pile-up dependence & $\pm 1\%$ & N/A \\
    MC event statistics & $\pm 1\%$ & N/A \\
    \hline
    Luminosity & \multicolumn{2}{c}{$\pm 3.2\%$} \\
    \bottomrule
  \end{tabular}
\end{table}

In order to facilitate the re-interpretation of the present results in alternative physics models,
the fiducial acceptance and efficiency for events 
with the invariant mass of the $\gamma$ + jet system around 
\mqstar or \mth 
(referred to as ``on-shell'' events hereafter) are also provided. 
The chosen \mgj ranges are $0.6\mqstar<\mgj<1.2\mqstar$ for the \qstar signal and $0.8\mth<\mgj<3.0\mth$ for the QBH signal.
The fiducial region at particle level,
as summarized in Table~\ref{tab:fiducial_definition}, is chosen to be close to the one used in the event selection
at reconstruction level. 
\begin{table}[htbp]
  \centering
  \caption{Requirements on the photon and jet at particle level to define the fiducial region and on the 
  detector-level quantities for the selection efficiency.}
  \label{tab:fiducial_definition}
    \begin{tabular}{c}
    \toprule
    Particle-level selection for fiducial region \\
    \midrule
    Photon : $\etgamma>150$~GeV, $|\etagamma|<1.37$ \\
    Jet : $\ptjet>60$~GeV, $|\Etajet|<4.5$ \\
    Photon--Jet $\eta$ separation : $|\detagj|<1.6$ \\
    No jet with $\ptjet>30$~GeV within $\DeltaR<0.8$ around the photon \\
    \midrule
    Detector-level selection for selection efficiency \\
    \midrule
    Tight photon identification \\
    Photon isolation \\
    Jet identification including quality and pile-up rejection requirements \\
    \bottomrule
  \end{tabular}
\end{table}

The fiducial acceptance $A_\mathrm{f}$, defined as the fraction of generated on-shell signal events falling into the fiducial region,
increases from 56\% to 63\% with increasing signal mass \mth from 1.0 to 6.5 (9.0)~TeV for the QBH in the RS1 (ADD) model.
The $A_\mathrm{f}$ value for the \qstar model varies very similarly to that for the RS1 QBH signal.
The rise in the fiducial acceptance as a function of \mth(\mqstar) is driven mainly by the increase of the efficiency
for the photon $\eta$ 
requirement since the photons tend to be more central as \mth(\mqstar) becomes larger.

\begin{figure}[htbp]
  \centering
  \subfloat[]{
    \includegraphics[width=0.45\textwidth]{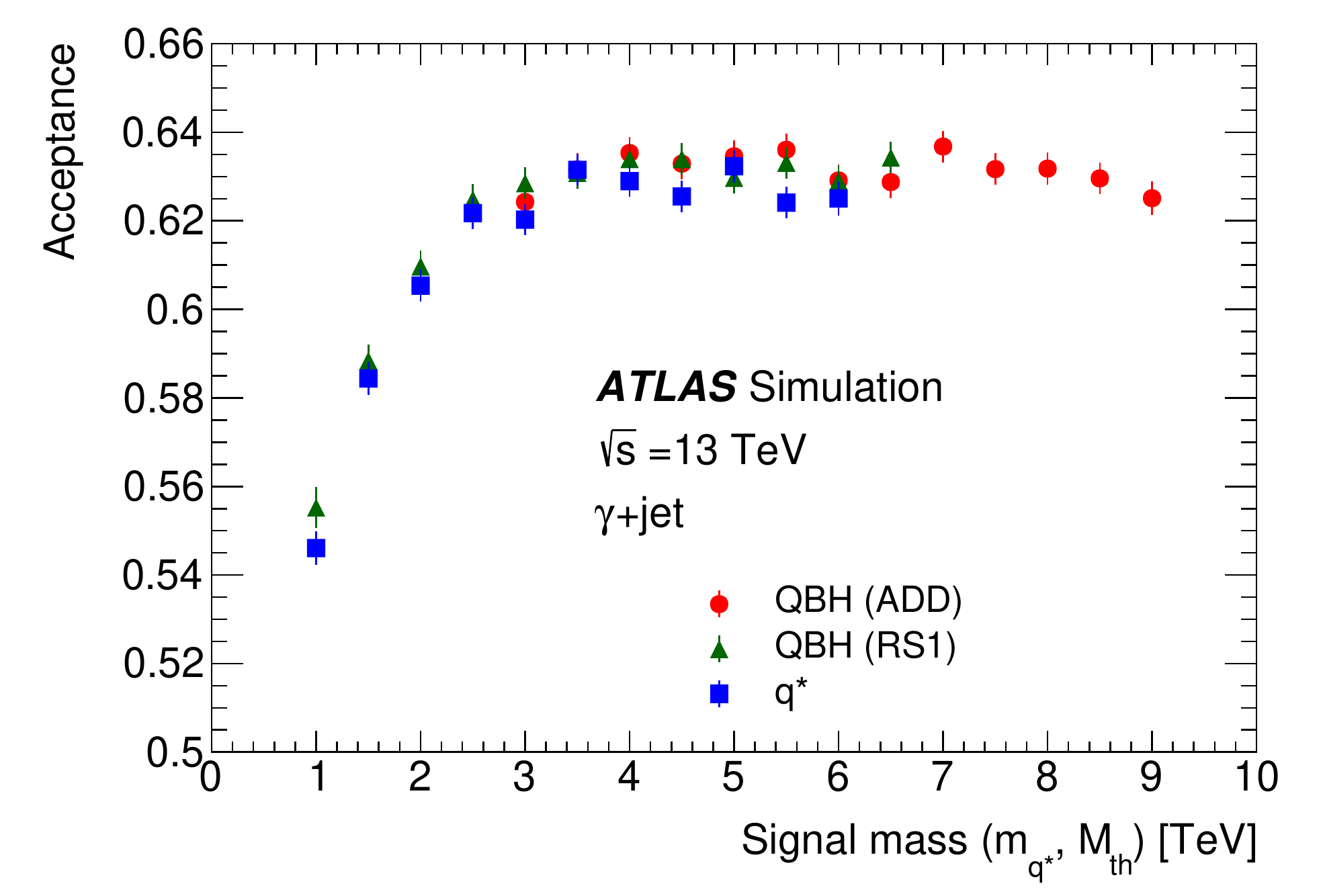}
    \label{fig:acceptance}
  }
  \subfloat[]{
    \includegraphics[width= 0.45\textwidth]{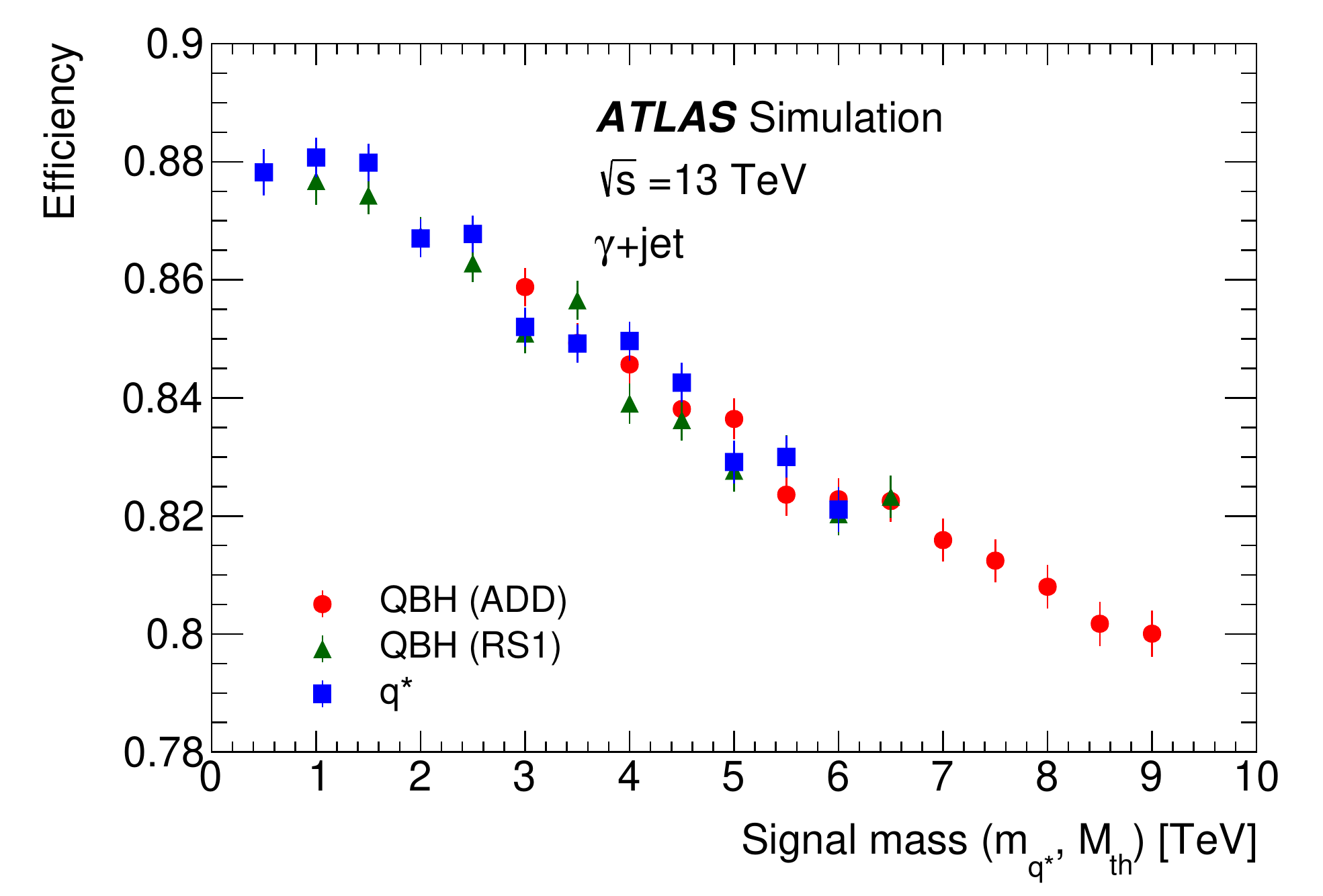}
    \label{fig:selection}
  }
  \caption{
    \protect\subref{fig:acceptance} Fiducial acceptance and \protect\subref{fig:selection} selection efficiencies 
    for the three signal models considered in the analysis 
    as a function of the excited-quark mass \mqstar or the QBH threshold mass \mth. 
    The fiducial region definition is detailed in Table~\ref{tab:fiducial_definition}.                                
    The description of the selection criteria can be found in the text.}          
  \label{fig:acc_eff}
\end{figure}

The fiducial selection efficiency $\varepsilon_\mathrm{f}$ is defined as the ratio of the number of on-shell events in the 
particle-level fiducial region passing the selection at the reconstruction level,
including photon identification, isolation and jet quality criteria, to the 
number of  generated on-shell events in the particle-level fiducial region.
The migration of generated on-shell events outside the particle-level fiducial region into the selected sample at the reconstruction level is found to be negligible.
The fiducial selection efficiency decreases from 88\% (86\%) to 82\% (80\%) within 
the same \mth ranges as above for the RS1 (ADD) QBH model
and is not highly dependent on the kinematics of the assumed signal production processes. 
The $\varepsilon_\mathrm{f}$ for the \qstar model behaves very similarly to that for the RS1 QBH model.
The reduction in the fiducial selection efficiency is caused mainly by the inefficiency of the shower shape 
requirements used in the photon identification for high-\ET photons. 
The fiducial acceptance and selection efficiencies for the three benchmark signal models are shown  
in Figure~\ref{fig:acc_eff} as functions of \mqstar or \mth. 

\subsection{Background modelling}
\label{sec:background_model}

The \mgj distribution of the background is modelled using a functional form. 
A family of functions, similar to the ones used in the previous searches for \gj 
~\cite{EXOT-2011-04,EXOT-2012-22,EXOT-2015-12} and 
\gg resonances~\cite{HIGG-2016-08} 
as well as dijet resonances~\cite{EXOT-2015-02} is considered:

\begin{equation}
f_{\text{bg}}(x) = N (1 - x)^{p} x^{\sum_{i=0}^k a_i (\log x)^i} \, ,
\label{eq:bck_func}
\end{equation}

where $x$ is defined as $\mgj/\sqrt{s}$,
$p$ and $a_{i}$ are free parameters, and $N$ is a normalization factor. The number of free parameters
describing the normalized mass distribution is thus $k+2$ with a fixed $N$, where $k$ is the stopping point
of the summation in Eq.~(\ref{eq:bck_func}).
The normalization $N$ as well as the shape parameters $p$ and $a_{i}$ are simultaneously determined 
by the final fit of the signal plus background model to data. 
The goodness of a given functional form in describing the background is quantified based on the
potential bias introduced in the fitted number of signal events. 
To quantify this bias the functional form under
test is used to perform a signal + background fit to a large sample of background events 
built from the \jetphox prediction. 
The large \jetphox event sample is used for this purpose as the shape of the background 
prediction can be extracted with sufficiently small statistical uncertainty.

\begin{figure}[htbp]
\centering
\includegraphics[width=0.65\textwidth]{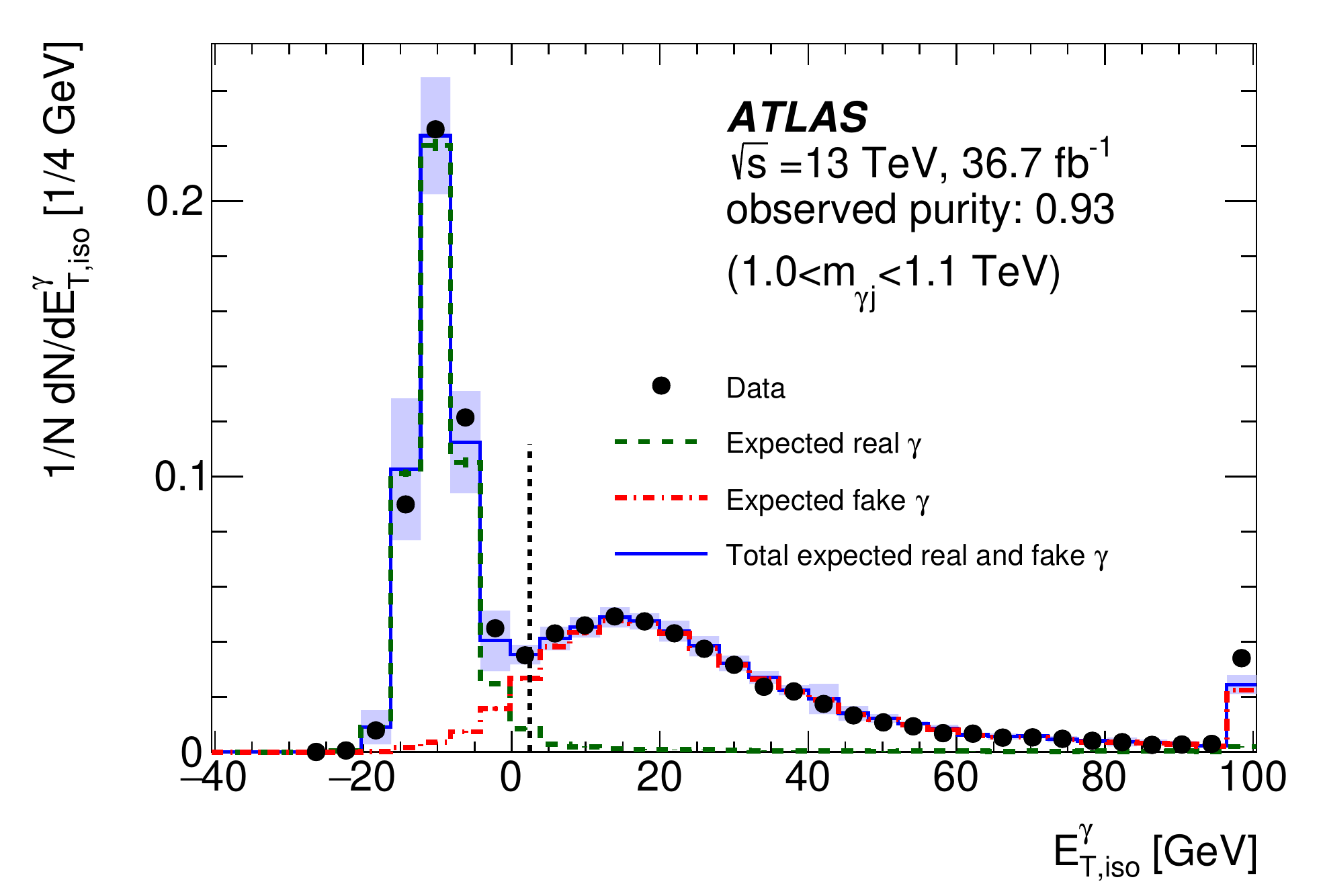} 
\caption{Distribution of $\etisogammaprime=\etiso-0.022\times\etgamma$ for the photon
candidates in events with $1.0<\mgj<1.1$~TeV, and the comparison with the result of the template fits. 
Real- and fake-photon components determined by the fit are shown by the green dashed and red dot-dashed histograms,
respectively, and the sum of the two components is shown as the solid blue histogram. 
The blue band shows the systematic uncertainties in the real- plus fake-photon template.
The last bin of 
the distribution includes
overflow events. The vertical dashed line corresponds to the isolation requirement used in the analysis.
The photon purity determined from the fit for the selected sample in the $1.0<\mgj<1.1$~TeV mass range is 93\%.}
  \label{fig:isolation}
\end{figure}

\begin{figure}[htbp]
\centering
\includegraphics[width=0.9\textwidth]{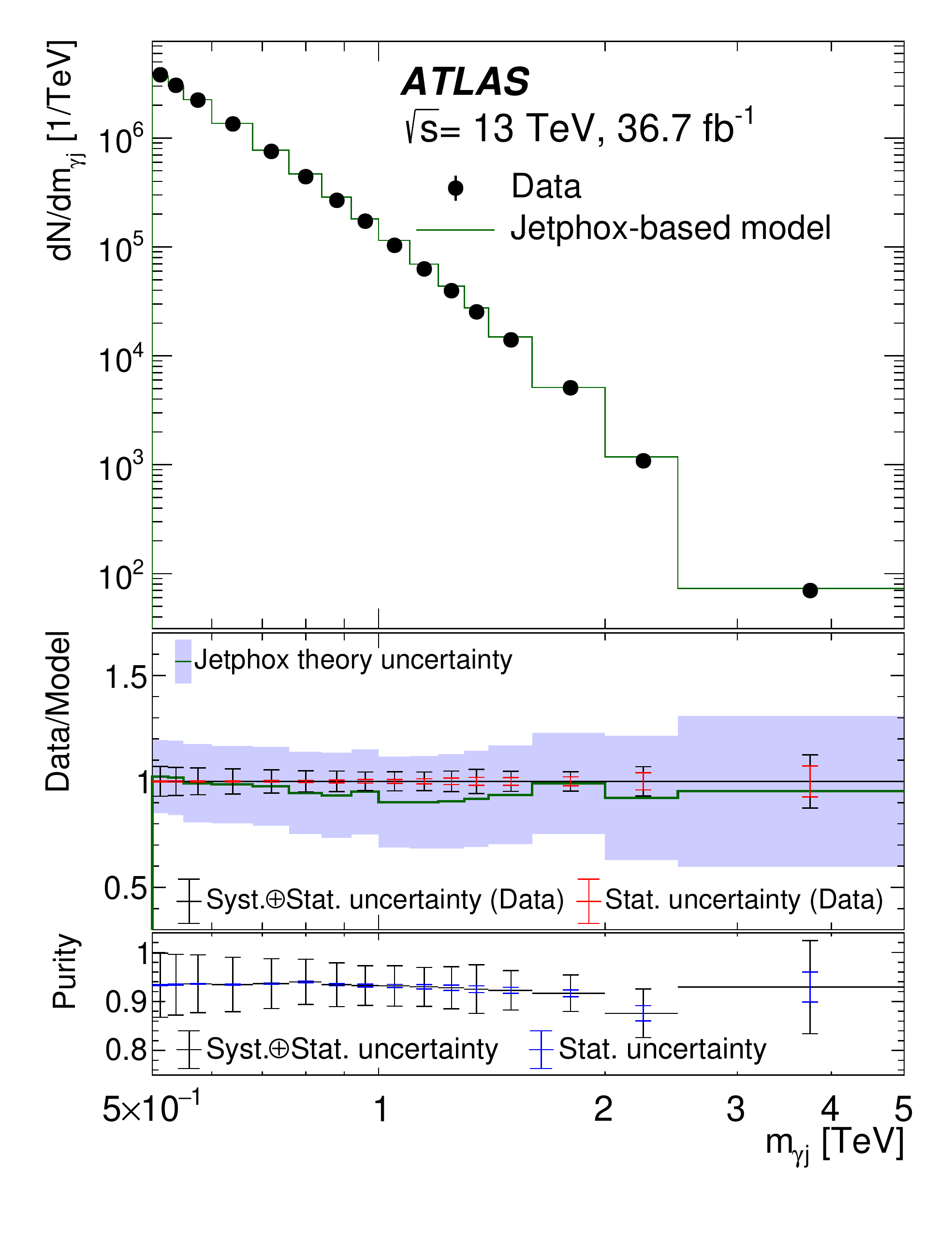}
\caption{Distribution of the invariant mass of the $\gamma$ + jet system as measured in the \gj data
(dots), compared with the \jetphox (green histogram) \gj predictions. 
The \jetphox distribution is obtained after correcting the parton-level spectrum for showering, hadronization
and detector resolution effects as described in the text. The distributions are divided by the bin width and 
the \jetphox spectrum is normalized to the data in the \mgj range above 500~GeV.
The ratio of the data to \jetphox prediction as a function of \mgj is shown in the middle panel
(green histogram): 
the theoretical uncertainty is shown as a shaded band. The statistical uncertainty from the data sample 
and the sum of the statistical uncertainty plus the systematic uncertainty from the background subtraction 
are shown as inner and outer bars respectively.
The measured \gj purity as a function of \mgj is presented in the bottom 
panel (black histogram): the statistical uncertainty of the purity measurement is reported as the 
inner error bar while
the total uncertainty is shown as the outer error bar.}
\label{fig:data_MC}
\end{figure}

The parton-level \jetphox calculations do not account for effects from hadronization, the underlying event and the detector resolution.
Therefore, the nominal \jetphox prediction 	
is corrected by calculating the ratio of reconstructed 
jet \pT to parton \pT in the \SHERPA \gj sample and applying the parameterized ratio to the \jetphox parton \pT.
In addition, an \mgj-dependent correction is applied to the \jetphox prediction to account for the 
contribution from multijet events where one of the jets is misidentified as a photon (fake photon events). 
This correction is estimated from data as the inverse of the purity, 
defined as the fraction of real \gj events in the selected sample.
The purity is measured in bins of \mgj by exploiting the difference 
between the shapes 
of the \etisogamma distributions of real photons and jets faking photons; the latter typically have a large	
\etisogamma value due to nearby particles produced in the jet fragmentation. The purity is estimated by performing a two-component 
template fit to the \etisogammaprime distribution in bins of \mgj.
The templates of real- and fake-photon isolation distributions are obtained from MC (\SHERPA) simulation and from data control samples, respectively. 
The \etisogammaprime variable for real photons from \SHERPA simulation is corrected to account for the 
observed mis-modelling in the description of isolation profiles between data and MC events in a separate control sample. 
The template for fake photons is derived in a data sample where the photon candidate fails to satisfy the tight 
identification criteria but fulfils a looser set of identification criteria. Details about 
the correction to the real-photon template and the derivation of the fake-photon template are given in Ref.~\cite{STDM-2015-11}.
To reduce the bias in the \etisogammaprime shape due to different kinematics, both the real-
and fake-photon templates are obtained by applying the same set of kinematic requirements used in the main analysis. 
As an example, Figure~\ref{fig:isolation} shows the \etisogammaprime distribution of events within the range $1.0<\mgj<1.1$~TeV, 
superimposed on the best-fit result. 
This procedure is repeated in every bin of the \mgj distribution and the resulting estimate
of the purity is shown as a function of \mgj in Figure~\ref{fig:data_MC}. 
The uncertainty in the measured purity includes both the statistical and systematic uncertainties. 
The latter are estimated by recomputing the purity using different data control samples for the fake-photon template or alternative
templates for real photons obtained from  \PYTHIA simulation or removing the data-to-MC corrections applied to \etisogammaprime in
the \SHERPA sample and by symmetrizing the variations. The variation from different data control samples 
for the fake-photon template has the largest effect on the purity (4\% at $1.0 < \mgj < 1.1 $ TeV).
The measured purity is approximately constant at 93\% over 
the \mgj range above 500~GeV, indicating that the fake-photon contribution does not depend significantly on 
\mgj .
Figure~\ref{fig:data_MC} shows the \mgj distribution in data compared to the 
corrected \jetphox \gj prediction normalized to data in the $\mgj > 500$ GeV region.

Theoretical uncertainties in the \jetphox prediction are computed by considering
the variations induced by $\pm 1\sigma$ of the NNPDF~2.3 PDF uncertainties,
by switching between the nominal NNPDF~2.3 and CT10 or MSTW2008 PDFs, by 
the variation of the value of the strong coupling constant by $\pm 0.002$ 
around the nominal value of 0.118 and by the variation of 
the renormalization, factorization and fragmentation 
scales between half and twice the photon transverse momentum.
The differences between data and the corrected \jetphox prediction shown in Figure~\ref{fig:data_MC} 
are well within the uncertainties associated with the perturbative QCD prediction.

The number of signal events extracted by the signal + background fit to
the pure background model described above is called the \emph{spurious signal} ~\cite{HIGG-2012-27} and
it is used to select the optimal functional form and the \mgj 
range of the fit. 
In order to account for the assumption 
that the corrected \jetphox prediction itself is a good
representation of the data, the fit is repeated 
on modified samples obtained by changing the nominal shape to account for several effects:
firstly, the nominal distribution is corrected to follow 
the envelope of the changes induced by $\pm 1\sigma$ variations of the NNPDF~2.3 PDF uncertainty,
the variations between the nominal NNPDF~2.3 and CT10 or MSTW2008 PDFs,
the variation of the value of the strong coupling constant by $\pm 0.002$ 
around the nominal value of 0.118 and the variation of 
the renormalization, factorization and fragmentation 
scales between half and twice the photon transverse momentum; secondly the corrections for
the hadronization, underlying event and detector effects are removed;
and finally the corrections for the photon purity are changed within their estimated uncertainty.
The largest absolute fitted signal from all variations of the nominal background sample discussed above is
taken to be the spurious signal.

The spurious signal is evaluated at a number of hypothetical masses over a large search range. 
It is required to be less than 40\% of the background's statistical uncertainty, 
as quantified by the statistical uncertainty of the fitted spurious signal, anywhere in the 
investigated search range. In this way the impact of the 
systematic uncertainties due to background modelling on the analysis sensitivity is expected to be 
subdominant with respect to the statistical uncertainty. 
Functional forms that cannot meet this requirement are rejected.  
For different signal models, the functional form and fit range are determined separately.
All considered functions with $k$ up to two (four parameters) are found to fulfil 
the spurious-signal requirement when fitting in the range $1.1<\mgj<6.0$~TeV for the \qstar signal and 
$1.5~(2.5)<\mgj<6.0~(8.0)$~TeV for the RS1 (ADD) QBH signal. 
To further consolidate the choice of nominal background functional form, an $F$-test~\cite{10.2307/2340521} 
is performed 
to determine if the change in the $\chi^2$ value obtained by fitting the \jetphox sample with
an additional parameter is significant. 
The test indicates that the $k=0$ (1) functional form with two (three) parameters 
can describe the present data sufficiently well over the entire fit range for the QBH (\qstar) signal search,
and there is no improvement by adding more parameters to the background fit function. 

Given the fit range determined by the spurious signal test, the search is performed for the \qstar (RS1 and ADD QBH) signal 
within the \mgj range above 1.5 (2.0 and 3.0)~TeV, to account for the width of the expected signal.
The estimated spurious signal for the selected functional form is converted into a spurious-signal 
cross-section (\sigmaspur), which is included as the uncertainty due to background modelling in the statistical analysis.
The spurious-signal cross-section, and the ratio of the spurious-signal cross-section to its uncertainty 
($\delta\sigma_\mathrm{spur}$) and to the signal cross-section (\sigmasig) for the three
benchmark models under investigation are given in Table~\ref{tab:spurious} in the different search ranges.
While both \sigmaspur and \sigmaspur/$\delta\sigma_\mathrm{spur}$ decrease                                                                           
with the hypothesized signal mass, the ratio \sigmaspur/\sigmasig increases with \mqstar or \mth, becoming as large as                                  
15\% in the case of excited quarks with \mqstar = 6 TeV.

\begin{table}[htbp]
  \centering
  \caption{Spurious-signal cross-sections (\sigmaspur), and the ratio of the spurious-signal cross-sections to their 
  uncertainties ($\delta\sigma_\mathrm{spur}$) and to the signal cross-sections (\sigmasig) for the three benchmark 
  models. The values of these quantities are given at the boundaries of the search range reported in the first row.
}
  \label{tab:spurious}
    \begin{tabular}{lcccccc}
    \toprule
     &  \multicolumn{2}{c}{\qstar} & \multicolumn{2}{c}{RS1 QBH} & \multicolumn{2}{c}{ADD QBH} \\
    \midrule
    Search boundaries [TeV]   & 1.5 & 6.0                & 2.0 & 6.0                & 3.0                & 8.0 \\
    \sigmaspur [fb]           & 3.9 & $1.1\times10^{-2}$ & 4.0 & $6.6\times10^{-4}$ & $8.7\times10^{-2}$ & $5.0\times10^{-5}$ \\
    \sigmaspur/  $\delta\sigma_\mathrm{spur}$ [\%]& 37 & 14 & 39&8 & 20&3 \\		
    \sigmaspur/\sigmasig [\%] & 0.16 & 15 & 1.0 & 7.5 & 0.0017 & 0.037 \\ 
 	
    \bottomrule
  \end{tabular}
\end{table}

A similar test is performed to determine the functional form and fit ranges for the Gaussian-shaped signal with a
15\% width. The test indicates that the same functional form and fit range as those used for the \qstar signal 
are optimal for a wide-width Gaussian signal. 
The same functional form and mass range is used for all the Gaussian signals.

\subsection{Statistical tests}
\label{sec:fitting}

A profile-likelihood-ratio test statistic 
is used to quantify the compatibility 
of the data and the SM background prediction, and to set limits on the presence of possible signal
contributions in the \mgj distribution. The likelihood function \lh is built from a Poisson probability for 
the numbers of observed events, $n$, and expected events, $N$, in the selected sample:
\begin{equation*}
\lh = \text{Pois}(n|N(\np)) \times \left(\prod_{i=1}^n f(\mgj^i,\np)\right) \times G(\np),
\end{equation*}
where $N(\np)$ is the expected number of candidates,
$f(\mgj^i, \np)$ is the 
value of the probability density function of the invariant mass distribution evaluated for each candidate event \emph{i}
and \np are nuisance parameters. 
The $G(\np)$ term collects the set
of constraints on the nuisance parameters associated with the systematic uncertainties in the signal yield,
in the spurious signal and in the resolution (only for Gaussian signals)  and it is
represented by normal distributions centred at zero and with unit variance.  

The pdf of the \mgj distribution is given as the normalized sum of the signal and background pdfs:
\begin{equation*}
 f(\mgj^i,\np) = \frac{1}{N} [\nsig(\np_{\mathrm{yield}}) f_{\text{sig}}(\mgj^i) + \nbg f_{\text{bg}}(\mgj^i,\np_{\mathrm{bkg}})],
\end{equation*}
where \fsignom and \fbgnom are the normalized signal and background \mgj distributions described in the previous sections.
The $\np_{\mathrm{yield}}$ are nuisance parameters associated with the signal yield uncertainties (constrained) while
$\np_{\mathrm{bkg}}$ are the nuisance parameters of the background shape (unconstrained).
The expected number of events $N$ is given by the sum of the expected numbers of signal events (\nsig) and 
background events (\nbg). The \nsig term can be expressed as 
\begin{equation*}
\nsig(\np_{\mathrm{yield}}) = \nsigmodel + \nsigspur = (\sigmasig \cdot B \cdot A \cdot \varepsilon \cdot F(\boldsymbol{\delta}_{\varepsilon},\boldsymbol{\theta}_{\varepsilon})+ \sigmaspur \cdot \npspur)  \times \Lint \times F(\delta_{\ensuremath{\mathcal{L}}},\theta_{\ensuremath{\mathcal{L}}}),
\label{eq:nsig}
\end{equation*}
where \sigmaspur and \npspur are the spurious-signal cross-section described in Section~\ref{sec:background_model} and its nuisance parameter
while \Lint and $F(\delta_{\ensuremath{\mathcal{L}}},\theta_{\ensuremath{\mathcal{L}}})$ are the integrated luminosity
and its uncertainty. Apart from the spurious signal, systematic uncertainties with an estimated size $\delta_{X}$
are incorporated into the likelihood by multiplying the relevant parameter of the statistical model by a factor
$F(\delta_{X},\theta_{X})=\mathrm{e}^{\delta_{X} \theta_{X}}$.
The parameter of interest in the fit to Gaussian-shaped resonances is the visible cross-section 
$\sigmasig \cdot B \cdot A \cdot \varepsilon$
while that in the fit to \qstar and QBH signals is $\sigmasig \cdot B$. For the latter case, the additional nuisance parameters
for the signal efficiency uncertainties $F(\boldsymbol{\delta}_{\varepsilon},\boldsymbol{\theta}_{\varepsilon})$ are included.

The significance of a possible deviation from the SM prediction is estimated by computing the \pzero value, 
defined 
as the probability of the background-only model to produce an excess at least as large as the one observed in data.
Upper limits are set at 95\% confidence level (CL) with a modified frequentist \CLs method 
on the visible cross-section 
($\sigmasig \cdot B \cdot A \cdot \varepsilon$) for the Gaussian-shaped resonances or 
on the signal cross-section times branching ratio 
($\sigmasig \cdot B$) for the \qstar and QBH signals
by identifying the value for which the \CLs value is equal to $0.05$.

\section{Results}
\label{sec:result}
\begin{figure}[htbp]
  \centering
  \subfloat[]{
    \includegraphics[width=0.5\textwidth]{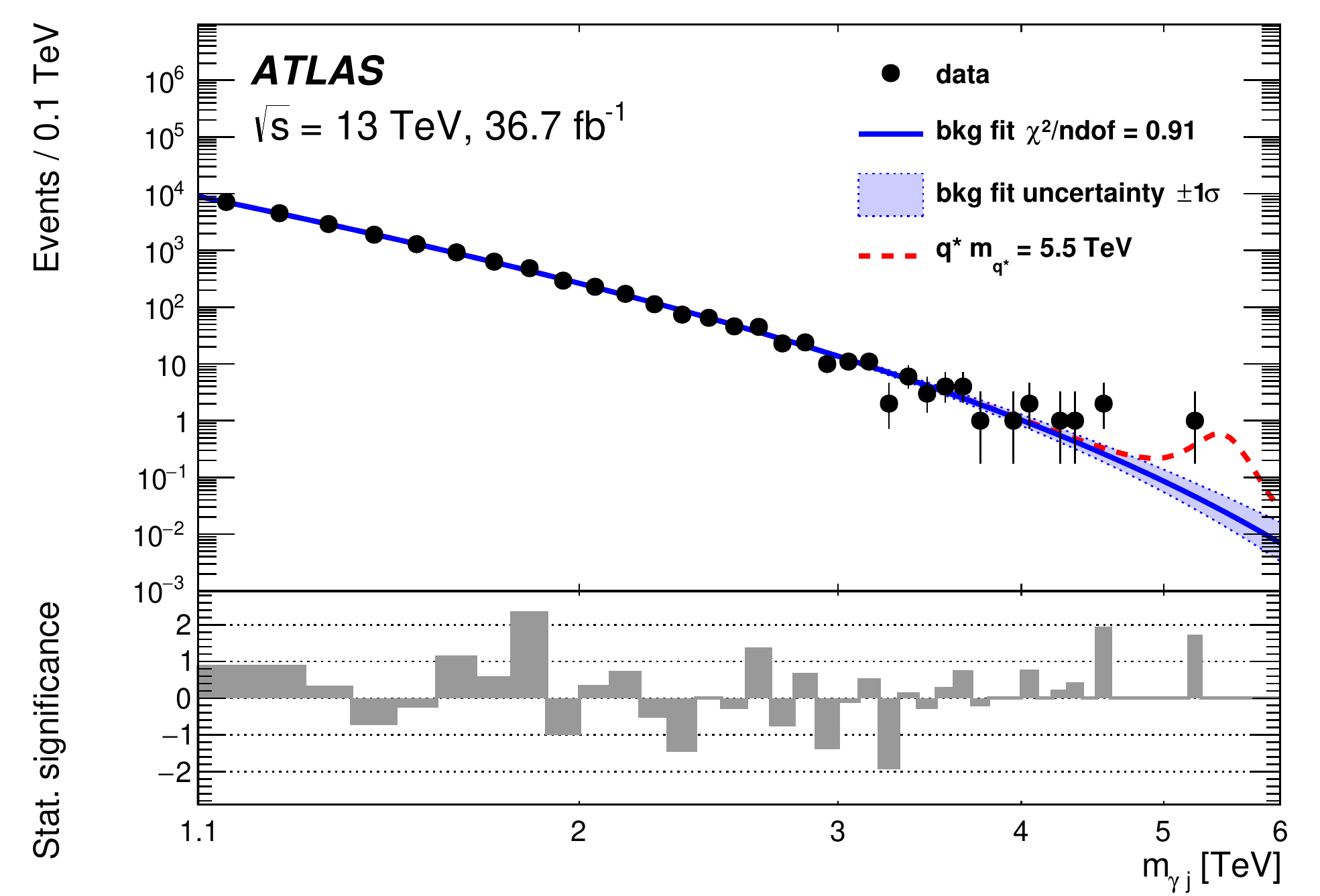}
    \label{fig:inv_mass_fit_1}
  }\\
  \subfloat[]{
    \includegraphics[width= 0.5\textwidth]{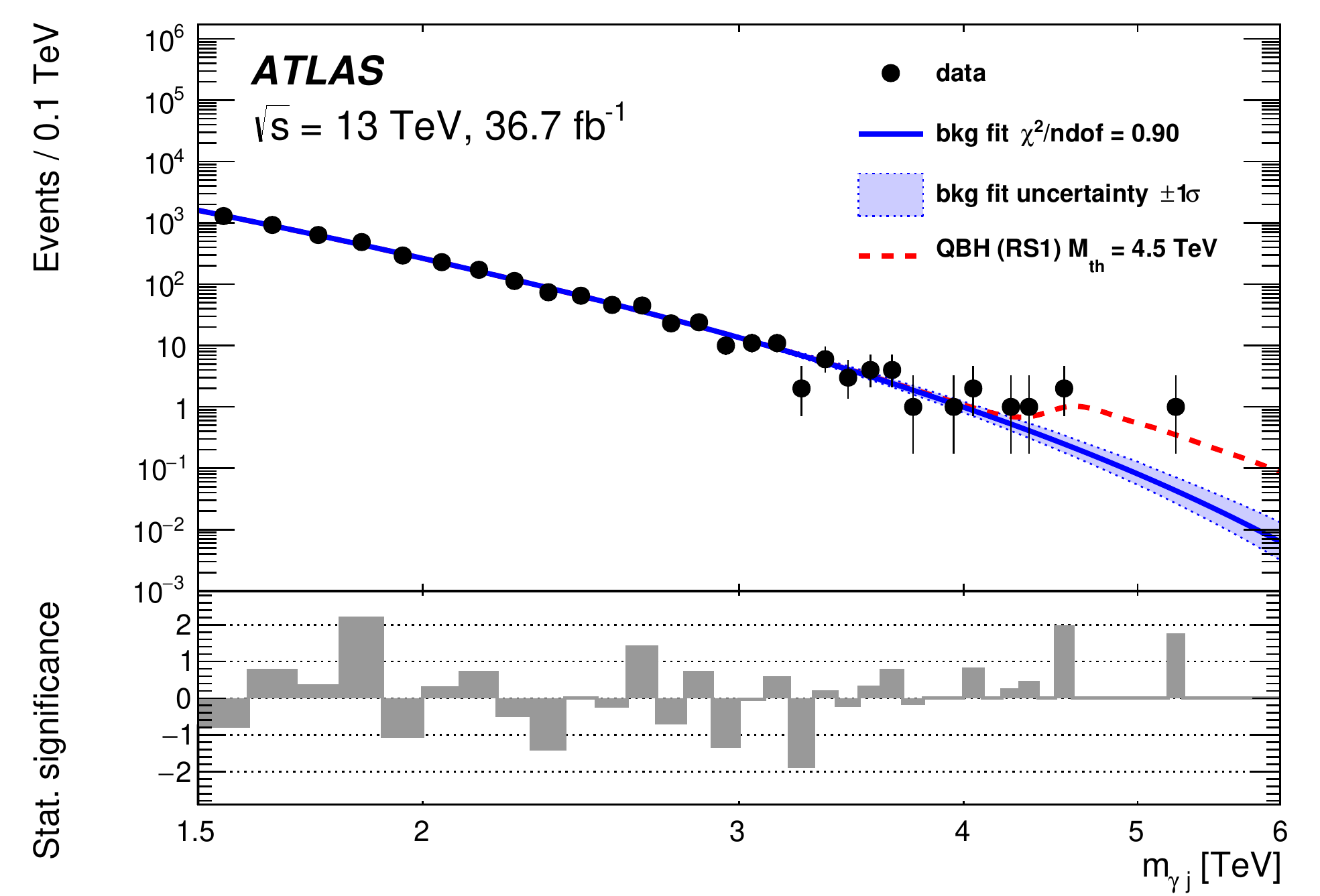}
    \label{fig:inv_mass_fit_2}
  }\\
  \subfloat[]{
    \includegraphics[width= 0.5\textwidth]{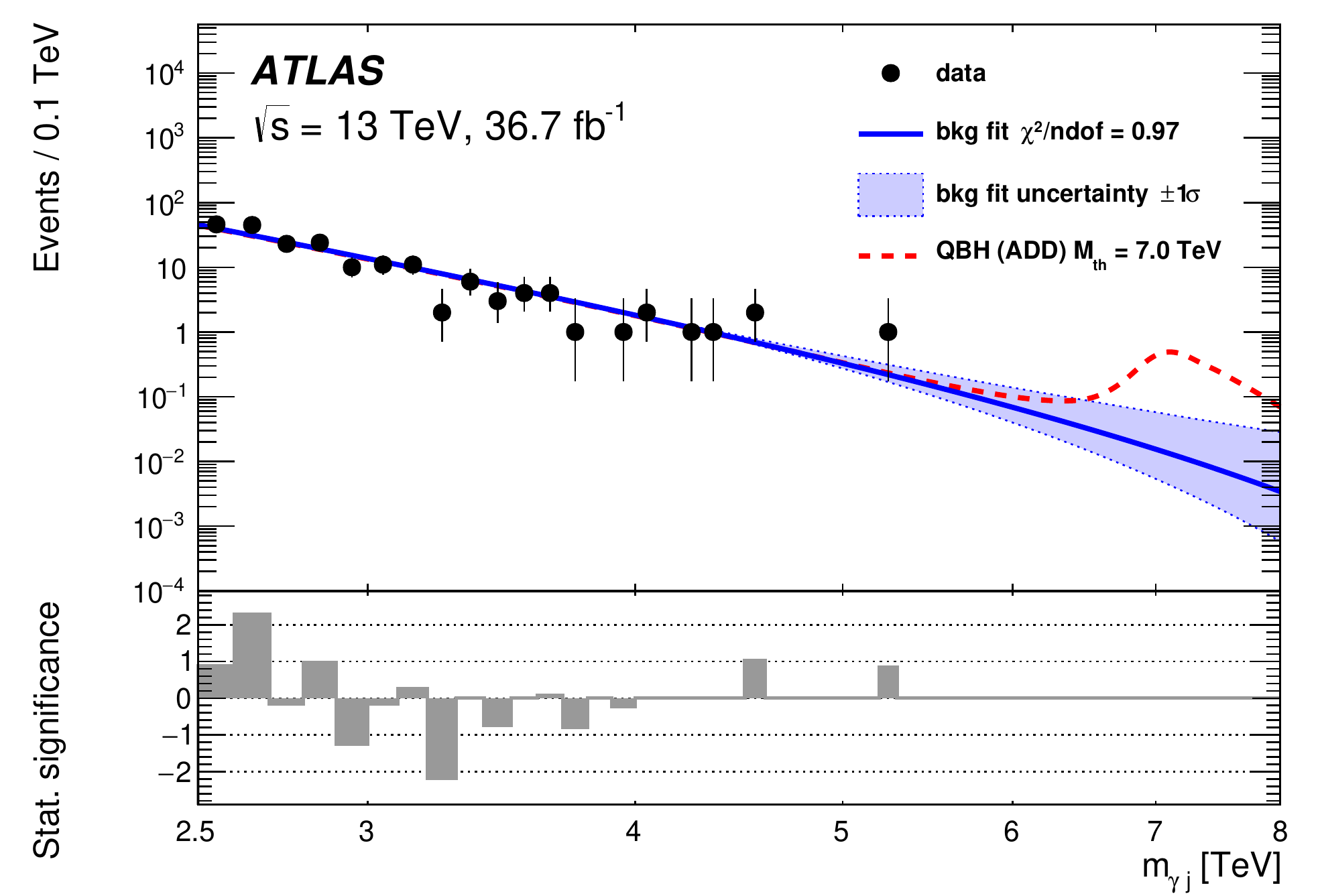}
    \label{fig:inv_mass_fit_3}
  }
  \caption{Distributions of the invariant mass of the $\gamma$ + jet system of the observed events (dots) in \lumi of data 
  at \sqs = 13~TeV and fits to the data (solid lines) under the background-only hypothesis for searches in the 
  \protect\subref{fig:inv_mass_fit_1} excited quarks,
  \protect\subref{fig:inv_mass_fit_2} QBH (RS1) with $n=1$ and \protect\subref{fig:inv_mass_fit_3} QBH (ADD) with $n=6$ models. 
  The $\pm1\sigma$ 
uncertainty in the background prediction originating from the uncertainties 
  in the fit function parameter values is shown as 
a shaded band around the fit.
  The predicted signal distributions (dashed lines)
  for the \qstar model with \mqstar = \mqstarAtXsecLimit~TeV and the QBH model with \mth = \mRSAtXsecLimit\xspace(\mADDAtXsecLimit)~TeV 
  based on RS1 (ADD) are shown on top of the background predictions. 
  The bottom panels show the bin-by-bin significances of the data--fit differences, considering only statistical uncertainties.
}          
  \label{fig:inv_mass_fit}
\end{figure}

The photon--jet invariant mass distributions obtained from the selected data are shown in Figure~\ref{fig:inv_mass_fit},
together with the background-only fits using the model described in Section~\ref{sec:background_model}
and expected distributions from the signal models under test.
No significant deviation from the background prediction is observed in any of the distributions.
The most significant excess is observed at 1.8~TeV with the assumption of the 2\%-width Gaussian model
for a local significance of 2.1 standard deviations. 

\begin{figure}[htbp]
  \centering  
  \includegraphics[width=0.7\textwidth]{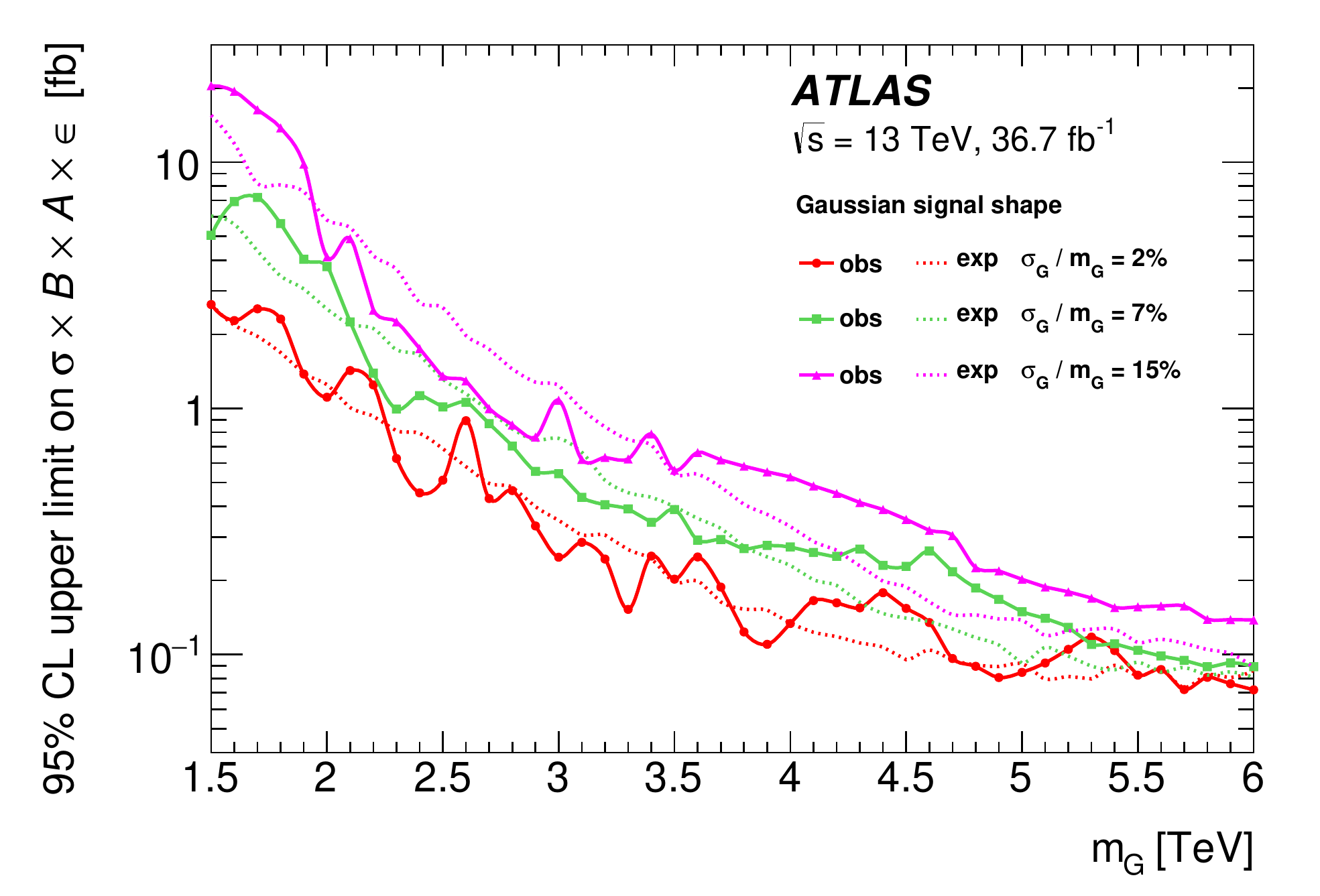}  
  \caption{Observed (solid lines) and expected (dotted lines) 95\% CL upper limits on the visible 
  cross-sections 
  $\sigma \cdot B \cdot A \cdot \varepsilon$ in \lumi of data at \sqs = 13~TeV
  as a function of the mass \mg of the Gaussian resonances with three different Gaussian widths between 2\% and 15\%.
  The calculation is performed using ensemble tests at mass points separated by 100~GeV over the search range.
  }        
  \label{fig:gaussian_limit}  
\end{figure}

\begin{figure}[htbp]
  \centering  
  \subfloat[]{
    \includegraphics[width=0.5\textwidth]{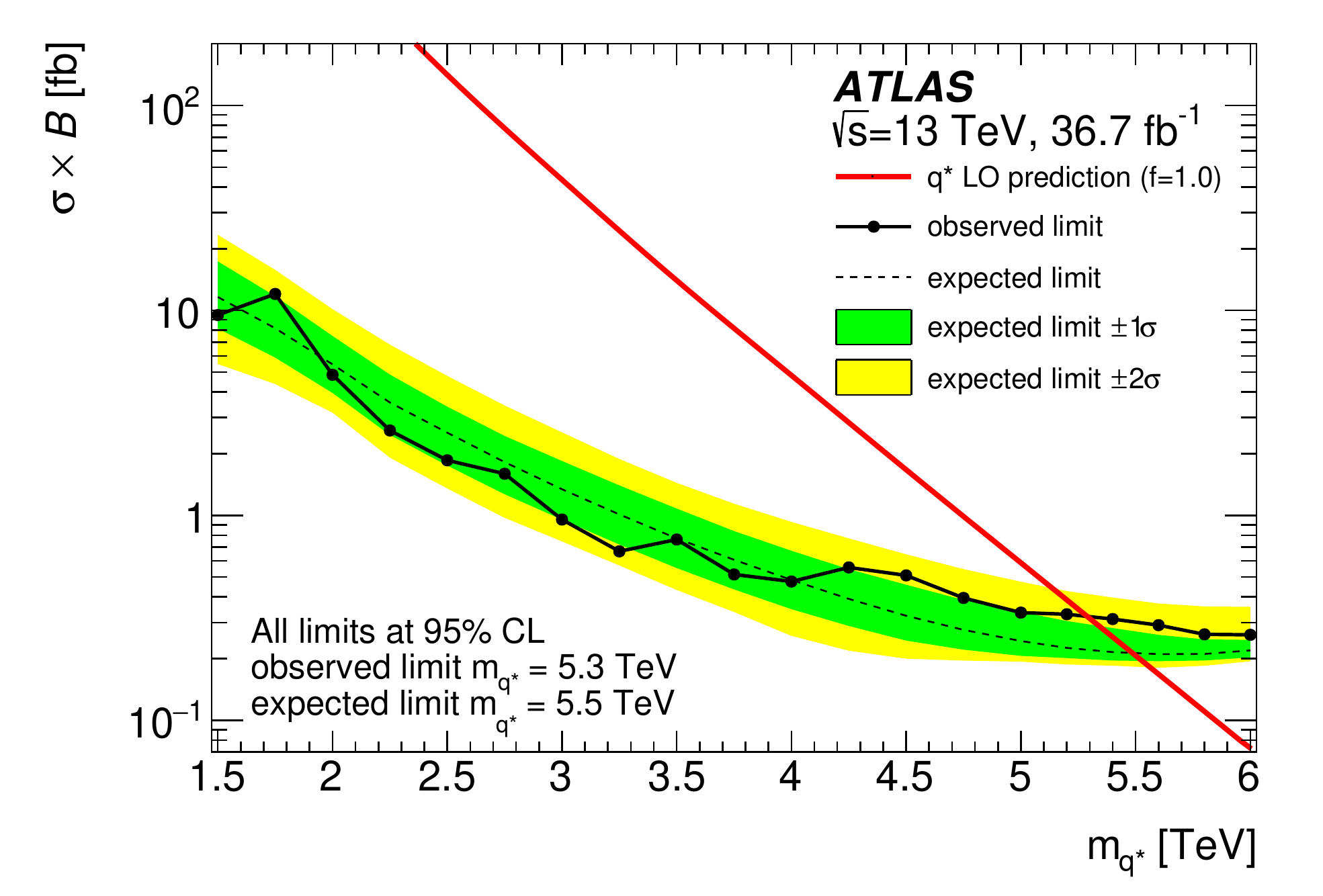}
    \label{fig:qstar_limit}
  }\\
  \subfloat[]{
    \includegraphics[width=0.5\textwidth]{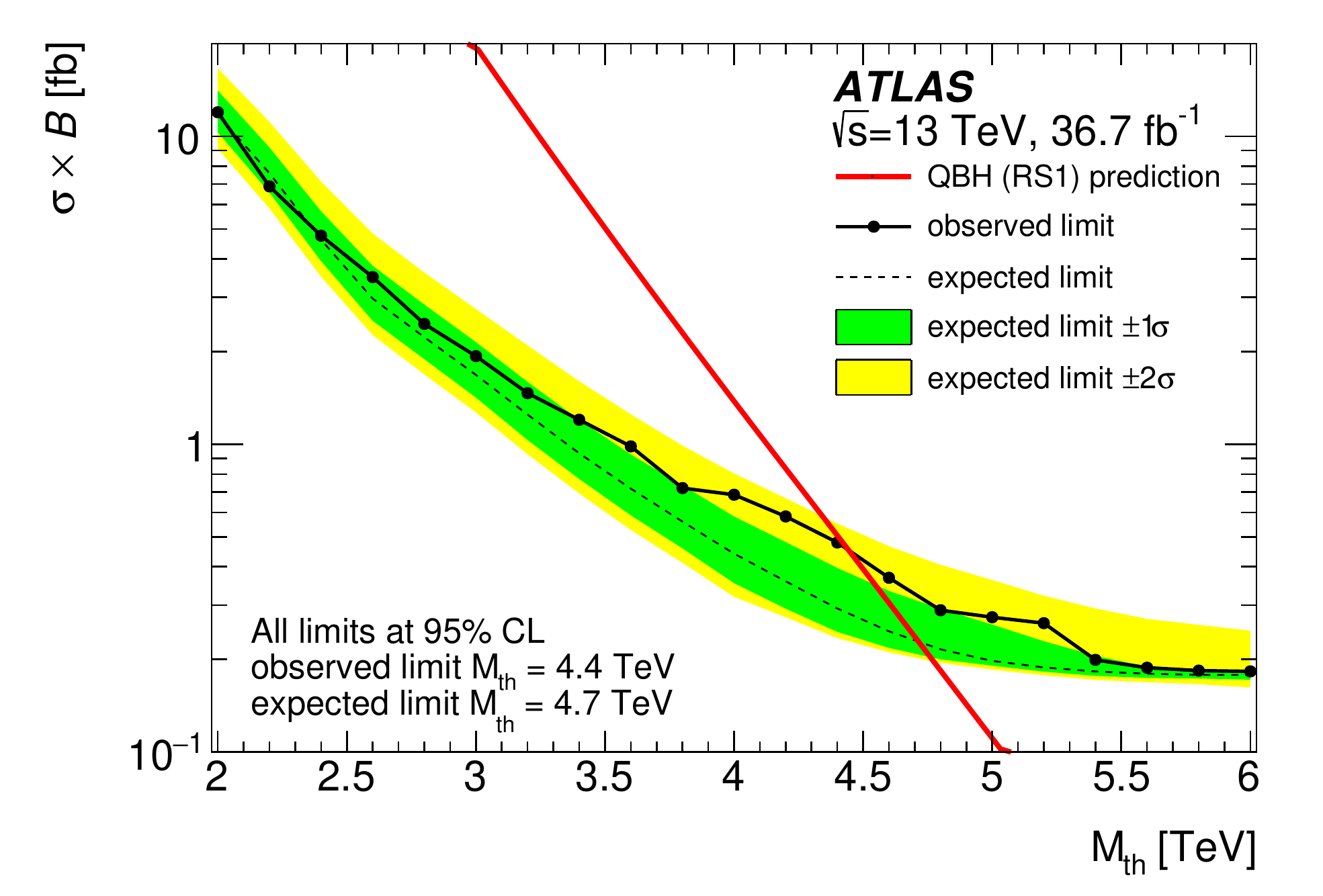}
    \label{fig:RS_limit}
  }\\
  \subfloat[]{
    \includegraphics[width=0.5\textwidth]{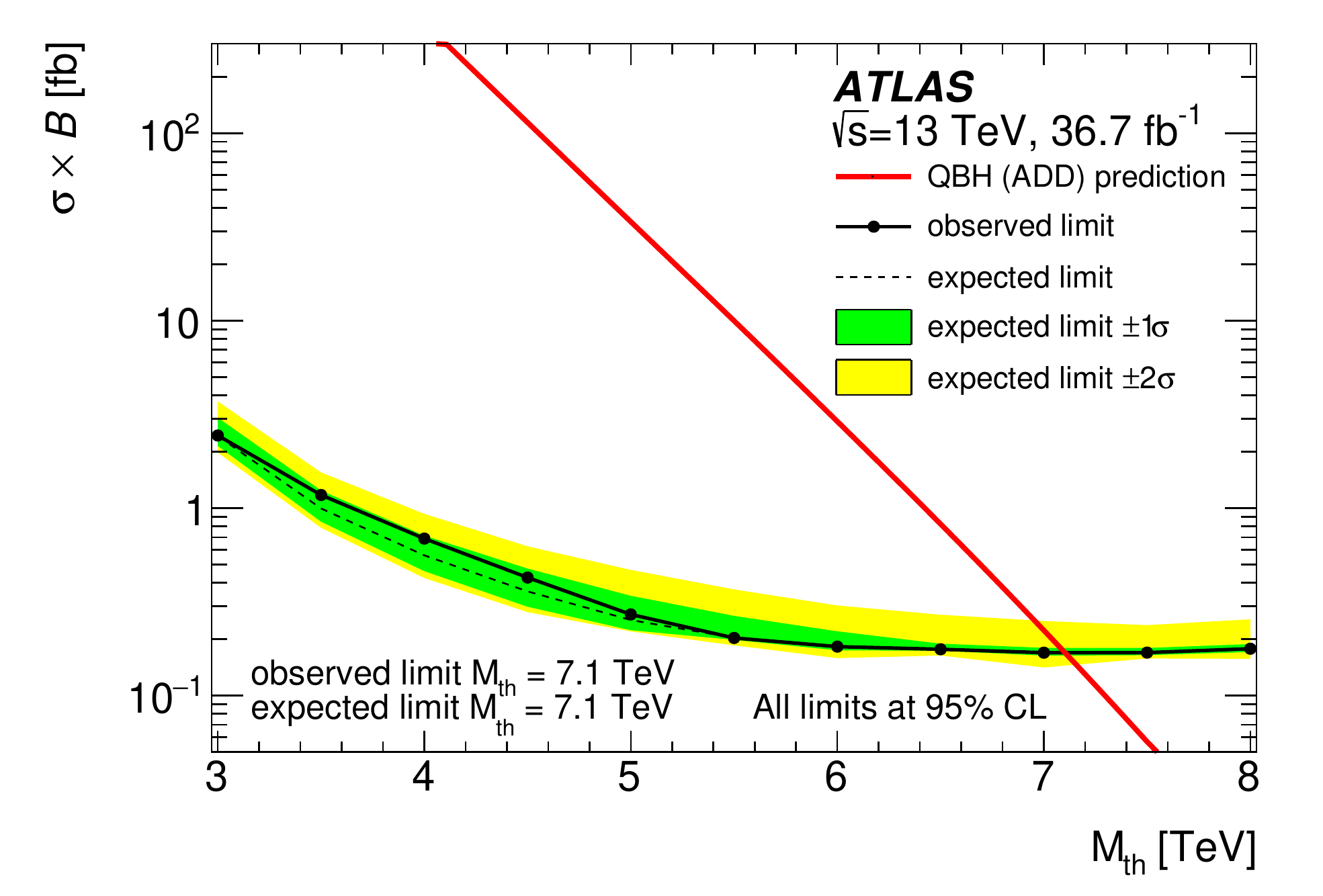}
    \label{fig:ADD_limit}
  }
  \caption{Observed 95\% CL upper limits (solid line with dots) on the production cross-section 
  times branching ratio
  \sigmabr to a photon and a quark or gluon in \lumi of data at \sqs = 13~TeV 
  for the \protect\subref{fig:qstar_limit} excited-quarks, 
  \protect\subref{fig:RS_limit} QBH (RS1) with $n=1$ and \protect\subref{fig:ADD_limit} QBH (ADD) with $n=6$ models. 
  The limits are placed as a function of \mqstar for the excited quarks and \mth for the QBH signals.
  The calculation is performed using ensemble tests at mass points separated by 200 (500)~GeV for the RS1 (ADD) model over the search range.
  For the \qstar model the step size is 250~GeV up to 5~TeV and then 200~GeV up to 6~TeV. 
  The limits expected if a signal is absent (dashed lines) are shown together with 
  the $\pm1\sigma$ and $\pm2\sigma$ intervals represented by the green and yellow bands, respectively. 
  The theoretical predictions of 
  \sigmabr for the respective benchmark signals are shown by the red solid lines. 
  }        
  \label{fig:model_limit}  
\end{figure}

Limits are placed at 95\% CL on the visible cross-section in the case of generic Gaussian-shaped resonances
and on the production cross-section times branching ratio 
to a photon and a quark or gluon for the excited-quark and 
QBH signals. The results are shown in Figure~\ref{fig:gaussian_limit} for the Gaussian signals with the width varying 
between 2\% and 15\%, and in Figure~\ref{fig:model_limit} for the benchmark signal models.
The Gaussian signals are excluded for visible cross-sections above 0.25--1.1~fb (0.08--0.2~fb), depending on the width, 
at a mass \mg of 3~TeV (5~TeV). 
In the case of the benchmark signal models considered in this analysis, 
the presence of a signal
with a mass below \mqstarObsLimit~TeV, \mRSObsLimit~TeV and \mADDObsLimit~TeV for the excited quarks, RS1 and ADD QBHs, 
can be excluded at 95\% CL.
The limits improve on those in Ref.~\cite{EXOT-2015-12} by about \mqstarObsLimitInc~TeV, 
\mRSObsLimitInc~TeV and \mADDObsLimitInc~TeV for the excited quarks, RS1 and ADD QBHs, respectively.

\FloatBarrier

\section{Conclusion}
\label{sec:conclusion}
A search is performed for new phenomena in events having a photon with high transverse momentum and a jet collected in 
\lumi of \pp collision data at a centre-of-mass energy of \sqs = 13~TeV recorded with the ATLAS detector at the LHC.
The invariant mass distribution of the $\gamma$ + jet system above 1.1~TeV is used in the search for localized 
excesses of events.
No significant deviation is found.
Limits are set on the visible cross-section for generic Gaussian-shaped resonances 
and on the production cross-section times branching ratio for signals predicted in models of excited quarks 
or quantum black holes. 
The data exclude, at 95\% CL, the mass range below \mqstarObsLimit~TeV for the excited quarks and 
\mADDObsLimit\xspace(\mRSObsLimit)~TeV 
for the quantum black holes with six (one) extra dimensions in the Arkani-Hamed--Dimopoulos--Dvali (Randall--Sundrum) model.
These limits supersede the previous ATLAS exclusion limits for excited quarks and quantum black holes in the
\gj final state.

\section*{Acknowledgements}

We thank CERN for the very successful operation of the LHC, as well as the
support staff from our institutions without whom ATLAS could not be
operated efficiently.

We acknowledge the support of ANPCyT, Argentina; YerPhI, Armenia; ARC, Australia; BMWFW and FWF, Austria; ANAS, Azerbaijan; SSTC, Belarus; CNPq and FAPESP, Brazil; NSERC, NRC and CFI, Canada; CERN; CONICYT, Chile; CAS, MOST and NSFC, China; COLCIENCIAS, Colombia; MSMT CR, MPO CR and VSC CR, Czech Republic; DNRF and DNSRC, Denmark; IN2P3-CNRS, CEA-DSM/IRFU, France; SRNSF, Georgia; BMBF, HGF, and MPG, Germany; GSRT, Greece; RGC, Hong Kong SAR, China; ISF, I-CORE and Benoziyo Center, Israel; INFN, Italy; MEXT and JSPS, Japan; CNRST, Morocco; NWO, Netherlands; RCN, Norway; MNiSW and NCN, Poland; FCT, Portugal; MNE/IFA, Romania; MES of Russia and NRC KI, Russian Federation; JINR; MESTD, Serbia; MSSR, Slovakia; ARRS and MIZ\v{S}, Slovenia; DST/NRF, South Africa; MINECO, Spain; SRC and Wallenberg Foundation, Sweden; SERI, SNSF and Cantons of Bern and Geneva, Switzerland; MOST, Taiwan; TAEK, Turkey; STFC, United Kingdom; DOE and NSF, United States of America. In addition, individual groups and members have received support from BCKDF, the Canada Council, CANARIE, CRC, Compute Canada, FQRNT, and the Ontario Innovation Trust, Canada; EPLANET, ERC, ERDF, FP7, Horizon 2020 and Marie Sk{\l}odowska-Curie Actions, European Union; Investissements d'Avenir Labex and Idex, ANR, R{\'e}gion Auvergne and Fondation Partager le Savoir, France; DFG and AvH Foundation, Germany; Herakleitos, Thales and Aristeia programmes co-financed by EU-ESF and the Greek NSRF; BSF, GIF and Minerva, Israel; BRF, Norway; CERCA Programme Generalitat de Catalunya, Generalitat Valenciana, Spain; the Royal Society and Leverhulme Trust, United Kingdom.

The crucial computing support from all WLCG partners is acknowledged gratefully, in particular from CERN, the ATLAS Tier-1 facilities at TRIUMF (Canada), NDGF (Denmark, Norway, Sweden), CC-IN2P3 (France), KIT/GridKA (Germany), INFN-CNAF (Italy), NL-T1 (Netherlands), PIC (Spain), ASGC (Taiwan), RAL (UK) and BNL (USA), the Tier-2 facilities worldwide and large non-WLCG resource providers. Major contributors of computing resources are listed in Ref.~\cite{ATL-GEN-PUB-2016-002}.

\clearpage
\printbibliography

\newpage 
\begin{flushleft}
{\Large The ATLAS Collaboration}

\bigskip

M.~Aaboud$^\textrm{\scriptsize 137d}$,
G.~Aad$^\textrm{\scriptsize 88}$,
B.~Abbott$^\textrm{\scriptsize 115}$,
O.~Abdinov$^\textrm{\scriptsize 12}$$^{,*}$,
B.~Abeloos$^\textrm{\scriptsize 119}$,
S.H.~Abidi$^\textrm{\scriptsize 161}$,
O.S.~AbouZeid$^\textrm{\scriptsize 139}$,
N.L.~Abraham$^\textrm{\scriptsize 151}$,
H.~Abramowicz$^\textrm{\scriptsize 155}$,
H.~Abreu$^\textrm{\scriptsize 154}$
% ...
\bigskip
\\
% ...
$^{12}$ Institute of Physics, Azerbaijan Academy of Sciences, Baku, Azerbaijan\\
% ...
$^{88}$ CPPM, Aix-Marseille Universit{\'e} and CNRS/IN2P3, Marseille, France\\
% ...
$^{115}$ Homer L. Dodge Department of Physics and Astronomy, University of Oklahoma, Norman OK, United States of America\\
% ...
$^{119}$ LAL, Univ. Paris-Sud, CNRS/IN2P3, Universit{\'e} Paris-Saclay, Orsay, France\\
% ...
$^{137}$ $^{(a)}$ Facult{\'e} des Sciences Ain Chock, R{\'e}seau Universitaire de Physique des Hautes Energies - Universit{\'e} Hassan II, Casablanca; $^{(b)}$ Centre National de l'Energie des Sciences Techniques Nucleaires, Rabat; $^{(c)}$ Facult{\'e} des Sciences Semlalia, Universit{\'e} Cadi Ayyad, LPHEA-Marrakech; $^{(d)}$ Facult{\'e} des Sciences, Universit{\'e} Mohamed Premier and LPTPM, Oujda; $^{(e)}$ Facult{\'e} des sciences, Universit{\'e} Mohammed V, Rabat, Morocco\\
% ...
$^{139}$ Santa Cruz Institute for Particle Physics, University of California Santa Cruz, Santa Cruz CA, United States of America\\
% ...
$^{151}$ Department of Physics and Astronomy, University of Sussex, Brighton, United Kingdom\\
% ...
$^{154}$ Department of Physics, Technion: Israel Institute of Technology, Haifa, Israel\\
$^{155}$ Raymond and Beverly Sackler School of Physics and Astronomy, Tel Aviv University, Tel Aviv, Israel\\
% ...
$^{161}$ Department of Physics, University of Toronto, Toronto ON, Canada\\
% ...
$^{a}$ Also at Department of Physics, King's College London, London, United Kingdom\\
$^{b}$ Also at Institute of Physics, Azerbaijan Academy of Sciences, Baku, Azerbaijan\\
$^{c}$ Also at Novosibirsk State University, Novosibirsk, Russia\\
$^{d}$ Also at TRIUMF, Vancouver BC, Canada\\
$^{e}$ Also at Department of Physics {\&} Astronomy, University of Louisville, Louisville, KY, United States of America\\
% ...
$^{*}$ Deceased
\end{flushleft}


\end{document}